\documentclass[12pt]{iopart}

\usepackage{graphicx}
\usepackage{dcolumn}
\usepackage{bm}
\begin{document}
\title{Bloch sphere like construction of SU(3) Hamiltonians using Unitary
Integration}
\author{Sai Vinjanampathy and A R P Rau}
\address{
Dept of Physics and Astronomy, Louisiana State University, Baton Rouge,
Louisiana 70803-4001,USA\\
}
\ead{saiv@phys.lsu.edu and arau@phys.lsu.edu}
\begin{abstract}
The Bloch sphere is a familiar and useful geometrical picture of the time
evolution of a single spin or a quantal  two-level system. The analogous
geometrical picture for three-level systems is presented, with several
applications. The relevant SU(3) group and su(3) algebra are eight-dimensional
objects and are realized in our picture as two four-dimensional manifolds that
describe the time evolution operator. The first, called the base manifold, is
the counterpart of the S$^2$ Bloch sphere, whereas the second, called the fiber,
generalizes the single U(1) phase of a single spin. Now four-dimensional, it
breaks down further into smaller objects depending on alternative
representations that we discuss. Geometrical phases are also developed and
presented for specific applications. Arbitrary time-dependent couplings between
three levels or between two spins (qubits) with SU(3) Hamiltonians can be
conveniently handled through these geometrical objects.
\end{abstract}
\submitto{\JPA}
\maketitle
\section{\label{intro}Introduction}
Three-level systems are of fundamental importance to many branches of
physics. While two levels give the simplest model for
the dynamics of discrete systems, three levels illustrate the role that an
intermediate state can play in inducing
transitions between the other two.
Canonical examples of this include applications in quantum optics
that use three-level atoms to control quantum state evolution
\cite{eit}. Such laser control is used, for instance, to
transfer population between two states using stimulated Raman adiabatic
passage (STIRAP)
\cite{stirap,stirap2} and chirped adiabatic passage (CARP) \cite{Chelkowski}. In
some of these systems, the interaction of the radiation with the atom is
represented as a time-dependent Hamiltonian inducing an energy
separation between the two states that varies with time. For a non-zero sweep
rate, it can be shown that there is  finite transition probability between the
states \cite{LZ,LZ2,LZ3}. The study of Landau-Zener transitions in multilevel
systems is
of interest to understand the interplay between various level
crossings \cite{vitanov}. Particle physics represents another example where
three-level
systems play a central role as, for example, the oscillations of neutrino flavor
eigenstates \cite{neutrino}.

The general Hamiltonian of a three-level system involves 8 independent
operators. Such a set can also naturally arise as a subgroup of higher level
systems where there is some degeneracy involved. Thus, several important
two-qubit
problems in quantum computing and quantum information can be so written in
terms of eight operators that
form a subalgebra of the full fifteen operators that describe two spins.  The
Hamiltonian describing anisotropic spin
exchange is an example of one such important physical problem. While
isotropic spin exchange has been explored to design two-qubit gates
in quantum computing, anisotropic spin exchange has been studied as
a possible impediment to two-qubit gate operations
\cite{divincenzo,divincenz2}. Such a SU(3) Hamiltonian is given by
\begin{eqnarray}\label{DV}
\mathbf{H}(t)=J(t)(\vec{\bm{\sigma}}.\vec{\bm{\tau}} +
\vec{\beta}(t).(\vec{\bm{\sigma}}
\times \vec{\bm{\tau}}) + \vec{\bm{\sigma}}.\mathbf{\Gamma}(t).\vec{\bm{\tau}}),
\end{eqnarray}
when written in terms of a scalar, a vector and a symmetric tensor operator
expressed in terms of two Pauli spins. Here, $\vec{\beta}(t)$ is the
Dzyaloshinksii-Moriya
vector \cite{dz1,dz2} and $\mathbf{\Gamma}(t)$ is the (traceless) symmetric
interaction term. While the first term is the familiar Ising interaction
Hamiltonian \cite{Chandler}, the last two terms are due to spin-orbit coupling.

\begin{figure}[!ht]
\centering
\includegraphics[height=2 in,width=3 in]{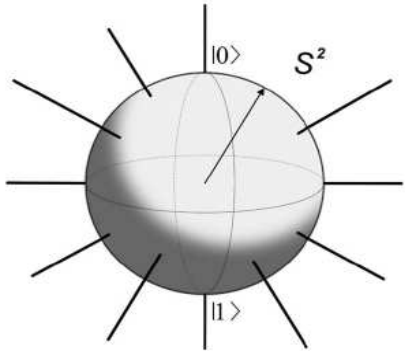}
\caption{Bloch or Poincare sphere representation for SU(2). The base
manifold is the $S^{2}$ sphere while the fiber is given by the U(1) phase at
each point on that sphere. Together, we have the fiber bundle
SU(2) $\simeq$ S$^2\times$U(1).}
\label{Fig1}
\end{figure}

 Given this wide applicability, a geometrical picture of the
dynamics of three-level systems can be useful. For a two-level system, the
geometry of the evolution
operator is well
known. Any density matrix can be written
as $\bm{\rho}=(I^{(2)}+\vec{n}.\vec{\bm{\sigma}})/2$, where $\vec{\bm{\sigma}}$
are the Pauli matrices. Unitary evolution of $\bm{\rho}$ is represented as the
vector $\vec{n}$  rotating on the surface of a three dimensional unit
sphere called the Bloch sphere \cite{poincare}. This vector, along with a
 phase, accounts for
the three parameters describing the time evolution operator of a two-level
system. The vector $\vec{n}$, along
with the phase factor, is shown in Fig. (\ref{Fig1}). The vector $\vec{n}$
shown traces
out the ``base manifold'' and together with the global phase factor or ``fiber''
at each point on that manifold is referred to as a ``fiber
bundle'' \cite{Bengtsson}.
While the
density matrix is independent of
it, the complete description of the system requires this phase as well. The aim
of this paper is to provide an analogous geometrical picture for a three-level
system with appropriate
generalizations of the base and fiber.

 Some work already exists regarding the geometry of SU(3). Following Wei
and Norman \cite{wei}, Dattoli and Torre have constructed the ``Rabi matrix"
for a general SU(3) unitary evolution in \cite{dattoli}. Mosseri and Dandoloff
in \cite{remy} described the generalization of the Bloch sphere
construction of single qubits to two qubits via the Hopf fibration
description. This method relies upon the homomorphism between the SU(2) and
SO(3) groups and likewise between the SU(4) and SO(6) groups. In \cite{tilma},
the authors propose a generalized Euler angle parameterization for SU(4). This
decomposition is similar to the work in
\cite{oldui,oldui2,oldui3,oldui4,restui,restui2,restui3} into which fits our
treatment of SU(3) in this paper.

Another well known choice of the $(N^{2}-1)$ generators $\mathbf{s}_{j}$ of the
SU(N) group was studied in \cite{hioe,dattoli2}. Consider $\mathbf{s}_{j}$,
chosen to be traceless and Hermitian such that
$[\mathbf{s}_{i},\mathbf{s}_{j}]=2i f_{ijk}\mathbf{s}_{k}$ and $Tr \{
\mathbf{s}_{i}\mathbf{s}_{j} \} = 2 \delta_{jk} $. Here, $f_{ijk}$ is the
completely antisymmetric symbol which for a two-level system is the Levi-Civita
symbol $\epsilon_{ijk}$, and a repeated index is summed over. In this basis, the
Hamiltonian is written as $
\mathbf{H}(t) = \Gamma_{i} \mathbf{s}_{i} $. With this choice, the Liouville-Von
Neumann equation for the density matrix
$\bm{\rho}=\mathbf{I}/N+S_{j}\mathbf{s}_{j}/2$ becomes
$\dot{S}_{i}=f_{ijk}\Gamma_{j}S_{k}$. Note that for the N=2 case, this is the
familiar Bloch sphere representation. But, for SU(3), this representation
differs from the one we
present in two aspects. Firstly, the ``coherence vector'', whose elements
are
real and are given by $S_{j}$, experiences rotations in a $(N^{2}-1)$
dimensional
space. For instance, for SU(3), the coherence vector undergoes rotations in an
eight-dimensional space. Arbitrary rotations in eight
dimensions are characterized by 28 parameters. But since a three-level
Hamiltonian is only characterized by 8 real quantities, this means that the
coherence vector is not permitted arbitrary rotations and is instead
constrained.
Secondly, the coherence vector representation does not differentiate between
local and non-local operations. Our decomposition of the time evolution operator
into a diagonal and an off-diagonal term in this paper is more suited for this
differentiation. Such a parameterization of the time evolution operator
in terms of local and non-local operations can be
useful in understanding entanglement. The aim of this paper is to
discuss the geometry of two-qubit time evolution operators in terms of such a
decomposition. The authors in \cite{englert} discuss an alternative
decomposition of two-qubit states in terms of two three-vectors and a $3 \times
3$ dyadic to discuss entanglement.

A series of papers presented a systematic approach to
studying N-level systems using a program of unitary integration
\cite{oldui,oldui2,oldui3,oldui4,newui,newui2,restui,restui2,restui3}.
Continuing this program, we
present a complete analytical solution to the three-level problem that
generalizes the Bloch sphere approach to three levels.
Below, we define the fiber bundle via two different decompositions which allows
us to extract the geometric
phases associated with a three-level system (for a discussion on the
quantum phases of three-level systems, see
\cite{gpqutrit,berry2and3}). These fiber bundles are $\{$SU(3)/SU(2)$\times$
U(1)$\}\times\{$SU(2)$\times$U(1)$\}$ and $\{$SU(4)/[SU(2)$\times$SU(2)]$\}
\times\{$SU(2)$\times$SU(2)$\}$.
The structure of this paper is as follows: Section \ref{ui} outlines
the unitary integration program to solve time-dependent operator
equations. Section \ref{su3} uses this technique for the
solution of a general time-dependent SU(3) Hamiltonian
completely analytically. Section
\ref{discussion} presents the geometry of the time
evolution operator for SU(3) with some applications. Section
\ref{GP} presents a coordinate description
that is useful to define the geometric phase for three-level systems, and
Section \ref{conclusions} presents the conclusions. The appendix will
present an alternative analytical solution to the three-level problem by
exploiting the natural embedding of SU(3) in SU(4).
\section{\label{ui}Unitary Integration}
Many important applications in physics involve time dependence in
the Hamiltonian. For such systems, the time evolution operator is
not given by the simple exponentiation of the Hamiltonian
\cite{sakurai}. To handle the time evolution for such Hamiltonians
iteratively, ``Unitary Integration" was proposed in
\cite{oldui,oldui2,oldui3,oldui4}.
Earlier work with this technique is presented in \cite{wei,dattoli2}. Later,
the technique was
presented as generalizing the SU(2) example
to solve iteratively for the time evolution operator $\mathbf{U}^{(N)}(t)$ of
N-level
systems \cite{newui,newui2}. Consider the N-dimensional Hamiltonian
$\mathbf{H}^{(N)}$
given by
\begin{equation}\label{Hamiltonian}
\mathbf{H}^{(N)}=\left(\begin{array}{cc}\mathbf{H}^{(N-n)}&\mathbf{V
}\\ \mathbf{V}^{\dagger}&
\mathbf{H}^{(n)}\end{array} \right).
\end{equation}
The diagonal blocks are (N$\--$n)- and (n)-dimensional square
matrices, respectively, while $\mathbf{V}$ is an $(N-n) \times (n)$-dimensional
matrix.

The evolution operator $\mathbf{U}^{(N)}(t)$ for such a
$\mathbf{H}^{(N)}$ is written as a product of
two operators
$\mathbf{U}^{(N)}(t)=\widetilde{U}_{1}\widetilde{U}_{2}$, where
\begin{eqnarray}\label{nonunit_u}
\widetilde{U}_{1}=\left(\begin{array}{cc}\mathbf{I}^{(N-n)}& \mathbf{z}(t)\\
\mathbf{0}^{\dagger}&
\mathbf{I}^{(n)} \end{array} \right)\left(\begin{array}{cc}\mathbf{I}^{(N-n)}&
\mathbf{0}\\
\mathbf{w}^{\dagger}(t)& \mathbf{I}^{(n)} \end{array} \right)
,\\
\widetilde{U}_{2}=\left(\begin{array}{cc}\widetilde{\mathbf{U}}^{(N-n)}&
\mathbf{0}\\ \mathbf{0}^{\dagger}&
\widetilde{\mathbf{U}}^{n} \end{array} \right).\nonumber
\end{eqnarray}
For any $N$, $n$ is arbitrary with $1 \leq n < N$, and tilde denotes
that the matrices need not be unitary. The product of three factors
parallels the product of exponentials in three Pauli matrices. Equations
defining
the rectangular matrices $\mathbf{z}(t)$ and $\mathbf{w}^{\dagger}(t)$ are
developed and the problem is reduced to the two residual $(N-n)$- and
$(n)$ dimensional evolution problems sitting as diagonal blocks of
$\widetilde{U}_{2}$. $\mathbf{z}(t)$ and
$\mathbf{w}^{\dagger}(t)$ are
related to each other through the unitarity of
$\mathbf{U}^{(N)}(t)$ \cite{newui,newui2}:
\begin{eqnarray}
\mathbf{z}=-\bm{\gamma}_{1}\mathbf{w}=-\mathbf{w}\bm{\gamma}_{2},
\end{eqnarray}
with $\bm{\gamma}_{1}=\hat{\mathbf{I}}^{(N-n)}+\mathbf{z}.\mathbf{z}^{\dagger}$
and
$\bm{\gamma}_{2}=\hat{\mathbf{I}}^{(n)}+\mathbf{z}^{\dagger}.\mathbf{z}$.

With $\mathbf{U}^{(N)}(t)$ in such a product form, the
Schr\"{o}dinger equation is written as
\begin{eqnarray}\label{effeqn}
i \dot{\widetilde{U}}_{2}(t)=\mathbf{H}_{\mathtt{eff}}\widetilde{U}_{2},\\
\nonumber
\mathbf{H}_{\mathtt{eff}}=\widetilde{U}^{-1}_{1}\mathbf{H}^{(N)}\widetilde{U}_{1
} \-- i \widetilde{U}^{-1}_{1}\dot{\widetilde{U}}_{1}.
\end{eqnarray}
Here, overdot denotes differentiation with respect to time. Since
$\widetilde{U}_{2}$ is block diagonal, the off-diagonal blocks of
equation~(\ref{effeqn}) define the equation satisfied by $\mathbf{z}$
\begin{equation}\label{zdot}
i\dot{\mathbf{z}}=\mathbf{H}^{(N-n)}\mathbf{z} + \mathbf{V}
\--\mathbf{z}(\mathbf{V}^{\dagger}\mathbf{z}+\mathbf{H}^{(n)}).
\end{equation}
Note that the initial condition $U^{N}(0)=\mathbf{I}^{N}$ implies
that $\widetilde{U}_{1}(0)=\mathbf{I}^{(N-n)}$,
$\widetilde{U}_{2}(0)=\mathbf{I}^{(n)}$
and $\mathbf{z}(0)=\mathbf{0}^{(N-n)}$. equation~(\ref{zdot}), along with
the initial condition can be solved to determine $\mathbf{z}$ and thereby
$\widetilde{U}_{1}$ and $\mathbf{H}_{\mathtt{eff}}$ for subsequent solution of
equation (\ref{effeqn}) for $\widetilde{U}_{2}$. In this manner, the
procedure iteratively determines $U^{(N)}(t)$.

Before discussing the geometry of the time evolution operators for
this unitary case, we briefly mention the procedure to deal with non-Hermitian
Hamiltonians. For such a non-Hermitian Hamiltonian,
\begin{equation}\label{Hamiltonian_non_Hermitian}
\mathbf{H}^{(N)}=\left(\begin{array}{cc}\widetilde{\mathbf{H}}^{(N-n)}&\mathbf{V
}\\ \mathbf{Y}^{\dagger}&
\widetilde{\mathbf{H}}^{(n)}\end{array} \right),
\end{equation}
where tilde denotes possibly non-Hermitian character, and the off-diagonal
components $\mathbf{V}$ and $\mathbf{Y}$ are independent.
In this case, equation (\ref{zdot}) is replaced by
\begin{equation}\label{zdot_non_Hermitian}
i\dot{\mathbf{z}}=\widetilde{\mathbf{H}}^{(N-n)}\mathbf{z} + \mathbf{V}
\--\mathbf{z}(\mathbf{Y}^{\dagger}\mathbf{z}+\widetilde{\mathbf{H}}^{(n)}),
\end{equation}
and there is a separate equation governing the evolution of $\mathbf{w}$ given
by
\begin{equation}\label{wdot_non_Hermitian}
i\dot{\mathbf{w}}^{\dagger}=\mathbf{w}^{\dagger}(\mathbf{z}\mathbf{Y}^{\dagger}
-\widetilde{\mathbf{H}}^{(N-n)})+(\widetilde{\mathbf{H}}^{(n)} +
\mathbf{Y}^{\dagger}\mathbf{z})\mathbf{w}^{\dagger} +\mathbf{Y}^{\dagger}.
\end{equation}
The diagonal terms of the time-evolution operators are governed by
\begin{eqnarray}\label{effeqn_non_Hermitian}
i \dot{\widetilde{U}}_{2}(t)=\left(\begin{array}{cc}
\widetilde{\mathbf{H}}^{(N-n)}-\mathbf{z}\mathbf{Y}^{\dagger}&0\\
0&\widetilde{\mathbf{H}}^{(n)}+\mathbf{Y}^{\dagger}\mathbf{z}
\end{array}
\right)\widetilde{U}_{2}.
\end{eqnarray}

Returning to the case where the Hamiltonian is Hermitian, it is convenient to
render the two matrices $\widetilde{U}_{1}$ and
$\widetilde{U}_{2}$  themselves unitary \cite{newui,newui2}. For this purpose, a
``gauge factor" $b$ is chosen such that the unitary counterparts of
$\widetilde{U}_{1}$ and $\widetilde{U}_{2}$ are defined via
$U_{1}=\widetilde{U}_{1}b$ and $U_{2}=b^{-1}\widetilde{U}_{1}$. Since
$\widetilde{U}^{\dagger}_{1}\widetilde{U}_{1}=diag(\gamma^{(-1)}_{1},\gamma_{2}
)$,
this would imply that b is the ``Hermitian square-root" of
$diag(\gamma^{(-1)}_{1},\gamma_{2})$. This ``Hermitian square-root"
is defined by the relation
$(b^{(-1)})^{\dagger}b^{(-1)}=diag(\gamma^{(-1)}_{1},\gamma_{2})$.
Inspection of the power series expansion of
$\gamma_{1}^{(\pm\frac{1}{2})}=(\hat{\mathbf{I}}+\mathbf{z}.\mathbf{z}^{\dagger}
)^{(\pm\frac{1}{2})}$
and
$\gamma_{2}^{(\pm\frac{1}{2})}=(\hat{\mathbf{I}}+\mathbf{z}^{\dagger}.\mathbf{z}
)^{(\pm\frac{1}{2})}$
show that since each term in the expansion is Hermitian, matrices
$\gamma^{\pm\frac{1}{2}}_{1}$ and $\gamma^{\pm\frac{1}{2}}_{2}$ are Hermitian
and
have non-negative eigenvalues. Because of this, it is
sufficient to define $b$ as the inverse square root via
$b^{(-2)}=diag(\gamma^{(-1)}_{1},\gamma_{2})$.

 Furthermore, $H_{\mathtt{eff}}$  in equation (\ref{effeqn}) is Hermitian for
the
unitary
counterpart $U_{1}$. The upper diagonal block of this  Hermitian
Hamiltonian accompanying the decomposition $U=U_{1}U_{2}$ is given
by
\begin{equation}\label{heffup}
\frac{i}{2}[\frac{d(\gamma_{1}^{-\frac{1}{2}})}{dt},\gamma_{1}^{\frac{1}{2}}]
+
\frac{1}{2}\left(\gamma_{1}^{-\frac{1}{2}}(\widetilde{\mathbf{H}}^{(N-n)}
-\mathbf { z }
\mathbf{V}^{\dagger})\gamma_{1}^{\frac{1}{2}}
+ H.c.\right),
\end{equation}
where [,] represents the commutator and \textit{H.c.} stands for the Hermitian
conjugate. The lower diagonal block is similarly given by
\begin{equation}\label{heffdown}
\frac{i}{2}[\frac{d(\gamma_{2}^{-\frac{1}{2}})}{dt},\gamma_{2}^{\frac{1}{2}}]
+
\frac{1}{2}\left(\gamma_{2}^{-\frac{1}{2}}(\widetilde{\mathbf{H}}^{(N-n)}
+\mathbf { z } ^
{\dagger}\mathbf{V})\gamma_{2}^{\frac{1}{2}}
+ H.c.\right).
\end{equation}
For $N=3$, $n=1$, these diagonal blocks define an SU(2)- and a U(1) Hamiltonian
and $\mathbf{z}$ is a pair of complex numbers.
The SU(2) Hamiltonian is in turn rendered in terms of its fiber bundle in Fig.
(\ref{Fig1}) and the U(1)
Hamiltonian corresponds to a phase. Together, they describe a four-dimensional
fiber for SU(3) over the base manifold, also four dimensional, of
$\mathbf{z}$.

Alternatively, $N=3$ SU(3) problems may be conveniently seen as a part of $N=4$
SU(4) problems, making contact with two qubit systems that are extensively
studied. In this case, for $N=4$, $n=2$, these diagonal blocks define two SU(2)
Hamiltonians and $\mathbf{z}$ is a $2\times2$ matrix representable in terms of
Pauli spinors. Generally, it is 8-dimensional while the fiber has seven
dimensions (two SU(2) and a mutual phase) but for the SU(3) subgroup of
SU(4),both the base and manifold again reduce to four dimensions each. With
$\textbf{z}$ a pair of complex numbers, the non-trivial part of geometrizing
SU(3) is thereby reduced to describing this four-dimensional manifold. Exploring
this for the $N=3$, $n=1$ decomposition will be the content of the next
 section whereas the Appendix gives the alternative SU(4) rendering.
\section{\label{su3}Geometry of general SU(3) time evolution operator}
A general time-dependent three-level Hamiltonian may be written in terms of
eight linearly independent operators of a three-level system. Such a Hamiltonian
can also be
written in terms of a subgroup of 15 operators of a four-level system. Before
the time evolution operator is presented in the SU(3) basis in terms of a $N=3$,
$n=1$ decomposition, we will note that it can be rendered in a few alternative
ways.

\begin{figure}[h!]
\centering
\includegraphics[height=2 in,width=2 in]{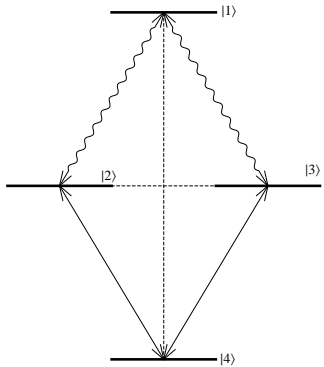}
\caption{Levels $\vert2\rangle$ and $\vert3\rangle$ couple equally to
$\vert1\rangle$ and to $\vert4\rangle$, which are themselves coupled. The three
complex coupling matrix elements and two energy positions define such an SU(3)
system.}
\label{Fig_energylevels}
\end{figure}

First, a general time-dependent four-level Hamiltonian may be written as
$H(t)=\sum_i c_{i}\mathbf{O}_{i}$. Here $c_{i}$ are time-dependent and
$\mathbf{O}_{i}$ are the unit matrix and 15 linearly independent
operators of a 4-level system that may be chosen in a variety of matrix
representations. One choice used in particle physics
are the so called Greiner matrices \cite{greiner,oldui,oldui2,oldui3,oldui4}.
Another choice
consists of using $\vec{\bm{\sigma}}$, $\vec{\bm{\tau}}$,
$\vec{\bm{\sigma}}\otimes\vec{\bm{\tau}}$ and the $4\times4$ unit matrix. Such
a choice was discussed in \cite{restui,restui2} and will be used throughout
this paper. As it stands, the above Hamiltonian describes a general
four-level atom with 4 energies and 6 complex couplings. Note that
only the three differences in energies are important. Restricting
the 15 coefficients $c_{i}$ to a smaller number allows this
Hamiltonian to describe various physical Hamiltonians, forming different
subalgebras of the su(4) algebra \cite{restui}. For
example, if two of the six complex couplings are zero (levels 1 and 4 and levels
2 and 3
of a four-level atom not coupled),
then the Hamiltonian may be recast such that the operators involved
belong to an so(5) subalgebra \cite{restui}. On the other hand, if levels 2
and 3
are degenerate and level 4 is uncoupled from the rest, then
the problem may be recast in terms of only eight operators belonging
to the su(3) subalgebra of su(4). This is illustrated in Fig.
\ref{Fig_energylevels} and is one of the systems of interest in this paper.

 Alternatively, after one arrives at the linear equation for the
$N=4$, $n=2$ decomposition, one can represent the resulting vector in terms of
six homogeneous coordinates. This is the so-called ``Pl\"{u}cker
coordinate'' representation for the SU(3) Hamiltonian. These coordinates as
well as the alternative
derivation are presented in the appendix. The $N=3$, $n=1$ decomposition will
be the content of the rest of this section.

Consider the Hamiltonian in the basis of the Gell-Mann lambda
matrices \cite{georgi} $H(t)=\sum_{i}a_{i}\bm{\lambda}_{i}$. The $N=3$, $n=1$
decomposition consists of writing the time evolution operator in terms of a
product of two matrices $U=\widetilde{U}_{1} \widetilde{U}_{2}$ where
$\widetilde{U}_{1}$ is composed of a (2$\times$1)-dimensional \textbf{z}, as
explained in Sec. II. The equation that governs the evolution of \textbf{z},
equation
(\ref{zdot}), can be written in this case as
\begin{equation}\label{su3_z_eqn}
\dot{z}_{\mu}=-iV_{\mu}-iF_{\mu\nu}z_{\nu}+iV^{*}_{\nu}z_{\nu}z_{\mu}
;\; \mu ,\nu=1,2.
\end{equation}
Here, the symbols used in defining $\dot{\mathbf{z}}$ are defined as
$V=(a_{4}-ia_{5},a_{6}-ia_{7})$,
and
\begin{equation*}
 F=\left(\begin{array}{cc}
 a_{3}+\sqrt{3} a_{8}&a_{1}-i a_{2}\\
a_{1}+i a_{2}&-a_{3}+\sqrt{3} a_{8}\\
\end{array} \right).
\end{equation*}
Using the transformation equations $m_{1,2}=-z_{1,2}(De^{i\phi})^{-1}$,
$m_{3}=(De^{i\phi})^{-1}$ and $\vert m_{1} \vert^{2}+\vert m_{2}\vert^{2}+\vert
m_{3} \vert^{2}=1$ leads to the evolution equation for
$\vec{m}=(m_{1r},m_{2r},m_{3r},m_{1i},m_{2i},m_{3i})^{T}$:
\begin{equation}\label{su3_Rotation}
\fl
\dot{\vec{m}}=\left(\begin{array}{cccccc}
0&-a_{2}&a_{5}&a_{3}+\sqrt{3}a_{8}&a_{1}&-a_{4}\\
a_{2}&0&a_{7}&a_{1}&-a_{3}+\sqrt{3}a_{8}&-a_{6}\\
-a_{5}&-a_{7}&0&-a_{4}&-a_{6}&0\\
-a_{3}-\sqrt{3}a_{8}&-a_{1}&a_{4}&0&-a_{2}&a_{5}\\
-a_{1}&a_{3}-\sqrt{3}a_{8}&a_{6}&a_{2}&0&a_{7}\\
a_{4}&a_{6}&0&-a_{5}&-a_{7}&0\\
\end{array} \right)
\vec{m},
\end{equation}
which describes the rotation of a unit vector in a six dimensional space of the
real and imaginary parts of $\vec{m}$ defined by $m_{\mu}=m_{\mu r}+im_{\mu
i}$. In the
above equations, $D=(1+\vert z_{1}\vert^{2}+\vert z_{2}\vert^{2})^{1/2}$
and $i\dot{\phi}=i(V^{*}_{\nu}z_{\nu}+V_{\nu}z^{*}_{\nu})$. The phase $\phi$ is
real and determined only up to a constant factor.
Since the real and imaginary parts of $m_{3}$ are not
independently defined, the geometrical
description of the base manifold  for the
$N=3$, $n=1$ decomposition may be
thought of as a point on the surface of a constrained six-dimensional
unit sphere.

\begin{figure}[!ht]
\centering
\includegraphics[height=2 in,width=4 in]{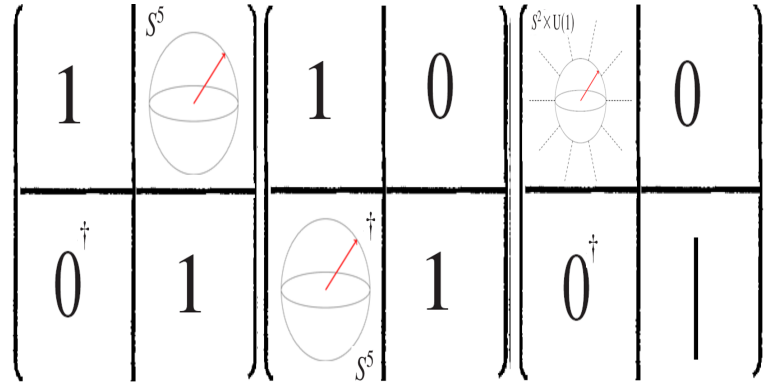}
\caption{The base and fiber for the SU(3) group. The first two factors give the
base manifold,
an $S^{5}$ sphere with a phase arbitrariness defined in the text. The fiber,
described by the third matrix, is
composed of a Bloch sphere and a phase
associated with each of its points, and the second an extra phase represented
by a vertical line.}
\label{Bloch3_SU3}
\end{figure}

The two constraints, namely $\vert m_{1}
\vert^{2}+\vert m_{2}
\vert^{2}+\vert m_{3} \vert^{2}=1$ and the ``phase arbitrariness" of
$\phi$, reduce the 6-dimensional manifold of the three-dimensional
complex vector $\vec{m}$ to a four-dimensional manifold in agreement with there
being only four independent parameters  in $\textbf{z}$.The first condition
defines the
base as a vector on an $S^{5}$ sphere while the phase arbitrariness serves as
an additional constraint. The fiber, on the other
hand, is an SU(2) block, evolving as a vector on  $S^{2}$
Poincare-like sphere with a phase at each point, and a U(1) block that amounts
to an extra phase.This is presented schematically in
Fig.(\ref{Bloch3_SU3}), as the product of three matrices of the evolution
operator.

\begin{figure}[!ht]
\centering
\includegraphics[height=2 in,width=4 in]{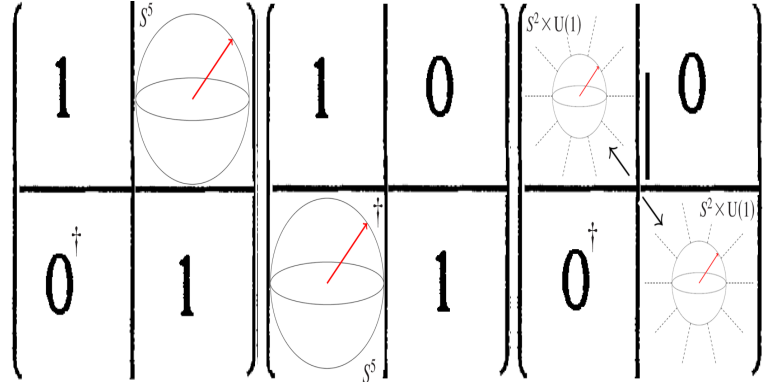}
\caption{The base and fiber for the SU(3) group via the $N=4$, $n=2$
decomposition. The base again is given by
an $S^{5}$ sphere as in Fig. (\ref{Bloch3_SU3}). The fiber is composed of two
identical SU(2) Bloch spheres plus phase, and an
extra mutual phase between them. The four parameters each of base and fiber
again account for all eight parameters
of the SU(3).}
\label{Bloch3_SU4}
\end{figure}

The alternative  $N=4$, $n=2$ decomposition in the appendix yields the
equation of motion for
$m_{\mu}=-z_{\mu}/De^{i \phi}$ in equation (\ref{rotation_SU4}).  Following
equation (\ref{heffup}) and equation (\ref{heffdown}), we see that for this
case, the two
remaining blocks of the time evolution operator, namely
$\widetilde{U}^{(4-2)}$ and $\widetilde{U}^{(2)}$, can be transformed
into unitary matrices for SU(2). The fiber evolves as vectors on two identical
$S^{2}$
Bloch-like spheres with a mutual phase, whose evolution is coupled to the base
that evolves as a
vector on an $S^{5}$ sphere. This is illustrated in Fig. (\ref{Bloch3_SU4}).
 Either decomposition can be used to study various physical processes as will
be discussed in the next section.

\section{\label{discussion}Applications}
\begin{figure}[!ht]
\centering
\includegraphics[height=1.7 in,width=2.75 in]{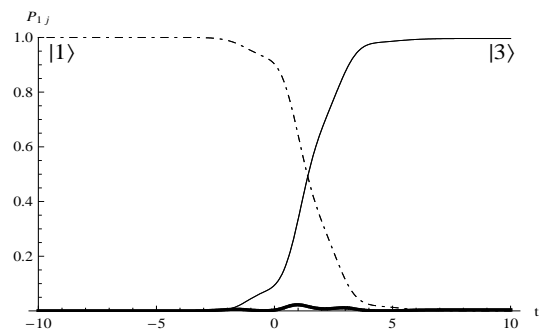}
\caption{Population $P_{1j}=\vert \langle1\vert j \rangle \vert^{2}$ plotted as
a function of time. The initial population in state $\vert 1 \rangle$
is  completely transferred to $\vert 3 \rangle$. Both the unitary
integration solution and the direct numerical solution \cite{mitra} are plotted
and they coincide at all times.}
\label{Fig_Mitra}
\end{figure}

It is often desirable to control the time evolution of quantum states to
manipulate an input state into a desirable output state. In \cite{mitra,mitra2},
the
authors considered a Hamiltonian of the form $\mathbf{H}_{0}-\mu
\mathcal{E}(t)$, where $\mathbf{H}_{0}$ is a free-field Hamiltonian and $\mu
\mathcal{E}(t)$ is a control field. To illustrate the ``Hamiltonian encoding''
scheme to control quantum systems, the authors considered a three-level system
and studied stimulated Raman adiabatic passage (STIRAP), an atomic
coherence effect that employs
interference between quantum states to transfer population completely from a
given initial state to a specific final state. This is done through a
``counterintuitive'' pulse sequence. Consider the Hamiltonian
\begin{eqnarray}\label{Ham_mitra}
 \mathbf{H}(t)=\left(\begin{array}{ccc}
0&G_{1}(t)&0\\
G_{1}(t)&2 \Delta&G_{2}(t)\\
0&G_{2}(t)&0
\end{array}\right).
\end{eqnarray}
Here $G_{1,2}(t)=2.5$exp$[-(t-t_{1,2})^{2}/\tau^{2}]$ and $\Delta=0.1$. The
initial
population is in the upper state. For $t_{1}=\tau$, $t_{2}=0$ and $\tau=3$, it
is seen that the two empty states are coupled first via $G_{2}(t)$ and then the
levels $\vert 1 \rangle$ and $\vert 2 \rangle$ are coupled through $G_{1}$. The
dynamics of the
populations reveal complete population transfer. A complete solution as per
Section \ref{su3} was constructed for this model and the results are presented
in Fig. \ref{Fig_Mitra} in total agreement with the results of \cite{mitra}.

\begin{figure}[!ht]
\centering
\includegraphics[height=4.8 in,width=3.2 in]{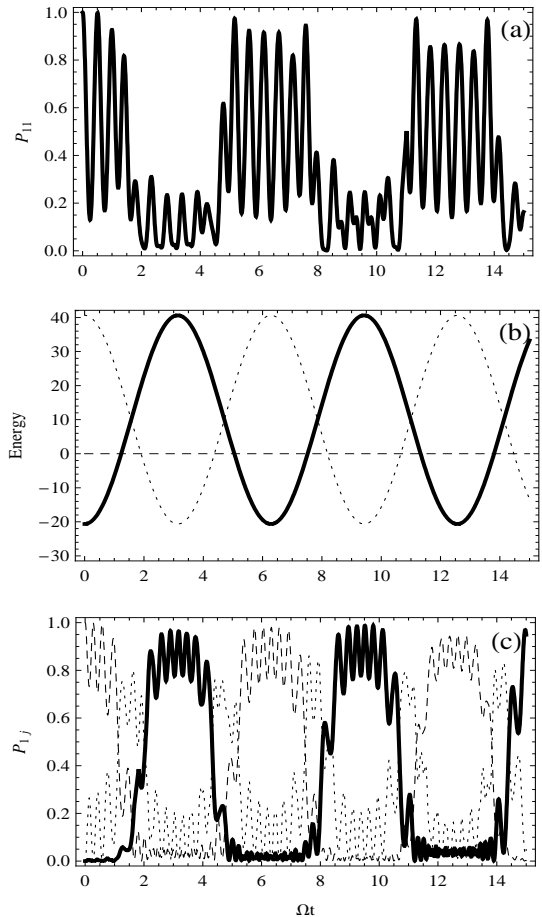}
\caption{(a) For $M_{1,2}=7$ and the other parameter values given in the text,
there is no population trapping observed. (b) The energy landscape for
$M_{1,2}=30.6346$ showing energy level crossing. (c) Population trapping is
observed with $M_{1,2}=30.6346$ which corresponds to the tenth zero of the
zeroth-order Bessel function. Note that the thick line is $P_{11}$ and the
thin line corresponds to $P_{12}$. The results agree completely with
\cite{agarwal}.}
\label{Fig_Agarwal}
\end{figure}

Quantum control can also be achieved by understanding the nature of tunneling.
The famous Landau-Zener formula \cite{LZ,LZ2,LZ3} predicts the transition
probability of
the ground state of a two-level system when the energy levels adiabatically
undergo a crossing. The study of level crossings has since been extended to
multi-level systems. For example, in \cite{agarwal}, the authors considered a
three-level atom to study population trapping by manipulating the phase acquired
as a three-level system evolves under the influence of frequency modulated
fields \cite{agarwal2}. Such a frequency modulated field is given by
\begin{eqnarray}
\mathbf{E}(t)=
\mathbf{E}_{1}e^{-i[\omega_{1}t+\varphi_{1}(t)]}+\mathbf{E}_{2}e^{-i[\omega_{2}
t+\varphi_{2}(t)]}+c.c.\\
\varphi_{i}(t)=M_{i}\sin{\Omega_{i}t}.
\end{eqnarray}
Here, $c.c.$ stands for complex conjugation. The phase $\varphi_{i}(t)$ in the
exponent can be written in terms of Bessel functions as \cite{Abramowitz}
\begin{equation}
 e^{M_{j}\sin{\Omega_{j}t}}=\sum^{\infty}_{k=-\infty}J_{k}(M_{j})e^{ik\Omega_{j}
t}.
\end{equation}
For large values of $\Omega_{j}$, the leading contribution for slow time
scales would come from $J_{0}(M_{j})$. Hence, for large $\Omega_{j}$, the
interaction Hamiltonian can be written as
\begin{eqnarray}
\mathbf{H}_{int}(t)=
-\mathbf{d}.(\mathbf{E}_{1}J_{0}(M_{1})+\mathbf{E}_{2}J_{0}(M_{2}).
\end{eqnarray}
Hence, for values of $M_{1,2}$ that are zeros of the zeroth-order Bessel
functions, the interaction Hamiltonian is zero and population trapping is
observed.
 Under this assumption, consider the  full Hamiltonian under the rotating wave
approximation,
\begin{eqnarray*}
 \mathbf{H}(t)=\left(\begin{array}{ccc}
E_{1}(t)&G_{1}(t)&0\\
G^{*}_{1}(t)&0&G_{2}(t)\\
0&G^{*}_{2}(t)&E_{3}(t)
\end{array}\right).
\end{eqnarray*}
Here, $E_{1}(t)=\Delta_{1}-M_{1}\Omega_{1}\cos(\Omega_{1} t+\theta)$ and
$E_{3}(t)=-\Delta_{2}+M_{2}\Omega_{2}\cos(\Omega_{2}t)$. Results are
presented in Fig. \ref{Fig_Agarwal}, and for the parameter
values $\Omega_{1,2}=1$, $\Delta_{1}=-\Delta_{2}=10$, $\theta=0$ and
$G_{1,2}=6$,
demonstrate the phenomenon of population localization discussed in
\cite{agarwal}.

\begin{figure}[!htbp]
\centering
\includegraphics[height=4.8 in,width=3.2 in]{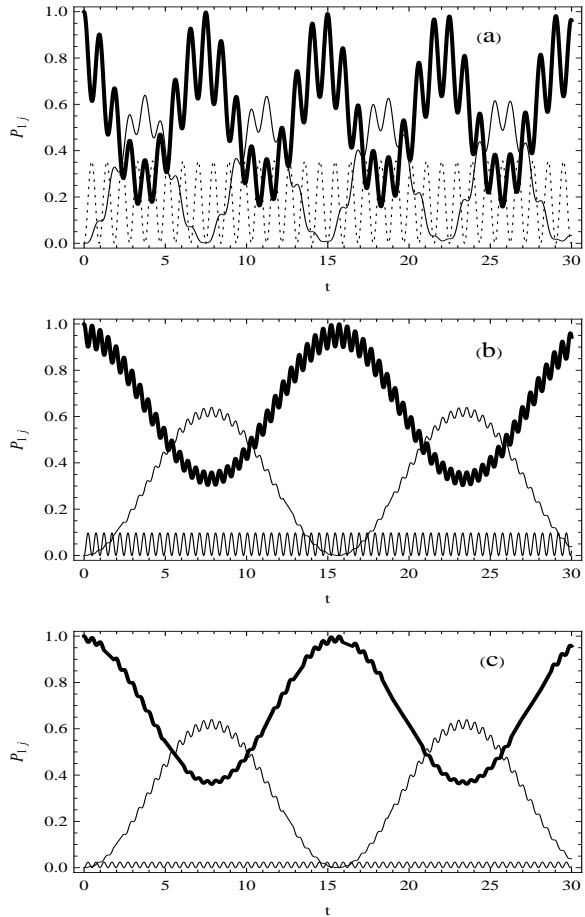}
\caption{(a) Populations $P_{1j}=\vert \langle 1 \vert j \rangle \vert^{2}$ for
$\delta=5$, $V_{1}=2$ and $V_{2}=1$. $P_{11}$ is given by the solid line and
$P_{12}$ is given by the thin line. (b)
Same as (a), for $\delta=12$. Note that $P_{13}$ oscillates close to zero at all
times. (c) $P_{1j}$ for $\delta=12$, $V_{1}=1$ and $V_{2}=2$.}
\label{Fig_kancheva}
\end{figure}

As a final illustration of the unitary integration technique applied to
three-level systems, let us consider the example discussed in \cite{kancheva}.
Here, a three-level system is subject to strong fields and the correlation
between the scattered light spectrum and the atom dynamics is discussed. The
authors consider the Hamiltonian
\begin{eqnarray}\label{Ham_kancheva}
 \mathbf{H}(t)=\left(\begin{array}{ccc}
0&0&G_{1}(t)\\
0&0&G_{2}(t)\\
G^{*}_{1}(t)&G^{*}_{2}(t)&0
\end{array}\right).
\end{eqnarray}
Here, $G_{1,2}(t)=-V_{1,2}e^{-i \delta t}$. The time evolution of the states
calculated as per our procedure in Section \ref{su3} is plotted in Fig.
\ref{Fig_kancheva} for different values of the parameters.
All of these results agree with those given in \cite{kancheva}. Further
features of the base and fiber will be presented at the end of the next section.

\section{\label{GP}Geometric phase for SU(3) group}
Many physical systems give rise to a measurable phase that does not depend
directly on the dynamical equations that govern the evolution of the system, but
depends only on the geometry of the path traversed by vectors characterizing the
state of the system. This geometric phase is denoted by $\gamma_{g}$ and is
given by the integral \cite{berry},
\begin{eqnarray}
 \gamma_{g}=\int d\mathbf{R}\;.\langle n(\mathbf{R}(t)) \vert
i\nabla_{\mathbf{R}} \vert n(\mathbf{R}(t)) \rangle,
\end{eqnarray}
where the state evolution is governed by a set of internal coordinates that
parameterize the Hamiltonian $\mathbf{R}(t)$, and $\nabla_{\mathbf{R}}$ is the
gradient in the space of these internal coordinates. This phase has been
generalized to non-cyclic non-adiabatic evolution of quantum
systems \cite{wilczek,wilczek2,wilczek3}. The purpose of this section is to
present
this phase in terms of coordinates on the Bloch sphere for two-level systems
and extend it to three-level systems.

In two-level systems, the time evolution operator is described by three
parameters as described in Section \ref{intro}. Two of these parameters describe
a point on the Bloch sphere. Traversing closed loops on this Bloch sphere
returns the
quantum system to its initial state as described
by the two parameters on the Bloch sphere but not the third parameter of an
overall phase.  Hence,
general closed loops on the Bloch sphere do not correspond to closed loops in
the space of the full unitary operator. This discrepancy in the phase between
the initial and final state corresponds to the geometric phase given above and
amounts to changes along the fiber at each point on the sphere. To
formalize this, consider $U_{1}$, given by equation (\ref{nonunit_u}) as
unitarized
through the matrix $b$ in Section \ref{ui}, which for $N=2$, $n=1$ takes the
form
\begin{eqnarray}
U_{1}=\frac{1}{\sqrt{1+\vert
\mathbf{z}\vert^{2}}}\left(\begin{array}{cc}1&\mathbf{z}\\ -\mathbf{z}^{*}&
1 \end{array} \right).
\end{eqnarray}

By identifying $\cos{\frac{\theta}{2}}=(1+\vert
\mathbf{z}\vert^{2})^{-\frac{1}{2}}$ and
$\sin{\frac{\theta}{2}}e^{-i \epsilon}=-\mathbf{z} (1+\vert
\mathbf{z}\vert^{2})^{-\frac{1}{2}}$, we get the usual description of the base
manifold in
terms of the angles $0\leq\theta<\pi$ and $0\leq\epsilon<2\pi$ that are
associated with the Bloch sphere, namely,
\begin{eqnarray}
U_{1}=\left(\begin{array}{cc}\cos{\frac{\theta}{2}}&-\sin{\frac{\theta}{2}}e^{
-i\epsilon}\\\sin{\frac{\theta}{2}}e^{i\epsilon}&
\cos{\frac{\theta}{2}} \end{array} \right).
\end{eqnarray}
In terms of the parameters $\theta$ and $\epsilon$, the Hamiltonian
$H(t)=-\vec{a}.\vec{\bm{\sigma}}$ is given by
\begin{eqnarray}
H(t)=\left(\begin{array}{cc}-\cos{\theta}&-\sin{\theta}e^{-i\epsilon}\\-\sin{
\theta}e^{i\epsilon}&
\cos{\theta} \end{array} \right).
\end{eqnarray}
equation (\ref{effeqn}) governing the evolution of the fiber $U_{2}$ has two
terms. The first term is evaluated as
\begin{eqnarray}
U^{\dagger}_{1}H(t)U_{1}=\left(\begin{array}{cc}-1&0\\0&
1 \end{array} \right),
\end{eqnarray}
which corresponds to the eigenvalues of the Hamiltonian. To evaluate the second
term, consider the case whereby the vector on the Bloch sphere traverses a
closed path defined by a constant $\theta$. The second term is then given by
\begin{eqnarray}
U^{\dagger}_{1}\frac{\partial{U}_{1}}{\partial(-i\epsilon)}=\left(\begin{array}{
cc}-\sin^{2}{\frac{\theta}{2}}&-\frac{1}{2}\sin{\theta}e^{-i\epsilon}\\-\frac{1}
{
2}\sin{\theta}e^{i\epsilon}&\sin^{2}{\frac{\theta}{2}}
\end{array} \right).
\end{eqnarray}
Integrating $\epsilon$ from 0 to $2\pi$ yields
\begin{eqnarray*}
\int^{2\pi}_{0}{d\epsilon}U^{\dagger}_{1}\frac{\partial{U}_{1}}{
\partial(-i\epsilon)}=\left(\begin{array}{cc}\pi(1-\cos{\theta}
)&0\\0&-\pi(1-\cos
{\theta})
\end{array} \right),
\end{eqnarray*}
which is the correct formula for the geometric phase of a two-level system
\cite{berry}.

To extend this analysis to three-level systems, we consider the $N=3$, $n=1$
decomposition. The matrix $U_{1}=\widetilde{U}_{1}.b$ is now given by
\begin{eqnarray}
U_{1}=\left(\begin{array}{cc}I^{(2)}-\frac{1}{D(D+1)}\mathbf{z}\mathbf{z}^{
\dagger}&\frac{\mathbf{z}}{D}\\-\frac{\mathbf{z}^{\dagger}}{D}&
\frac{1}{D} \end{array} \right),
\end{eqnarray}
where $\mathbf{z}$ is a complex column vector $(z_{1},z_{2})^{T}$ and
$D=\sqrt{1+\vert\mathbf{z}\vert^{2}}$. Care has to be taken in assigning
angles to elements of this matrix such that the transformation satisfies two
conditions: the $U_{1}$ matrix should not depend on $\phi$ and the
transformation must be
commensurate with the definition of $\vec{m}$. To this effect, we transform
$\mathbf{z}$ into polar coordinates:
$z_{1}=-\tan{\frac{\theta_{1}}{2}}\cos{\frac{\theta_{2}}{2}}e^{i\epsilon_{1}}$,
$z_{2}=-\tan{\frac{\theta_{1}}{2}}\sin{\frac{\theta_{2}}{2}}e^{i\epsilon_{2}}$.
These transformation equations imply that
$D=\sqrt{1+\vert\mathbf{z}\vert^{2}}=\sec{\frac{\theta_{1}}{2}}$,
$m_{1}=\sin{\frac{\theta_{1}}{2}}\cos{\frac{\theta_{2}}{2}}e^{i(\epsilon_{1}
-\phi)}$,
$m_{2}=\sin{\frac{\theta_{1}}{2}}\sin{\frac{\theta_{2}}{2}}e^{i(\epsilon_{2}
-\phi)}$ and $m_{3}=\cos{\frac{\theta_{2}}{2}}e^{-i\phi}$. The $U_{1}$
matrix is given by
%\begin{widetext}
\begin{eqnarray}
\fl
U_{1}=\left(\begin{array}{ccc}
1-2\sin^{2}{\frac{\theta_{1}}{4}}\cos^{2}{\frac{\theta_{2}}{2}}
&-\sin^{2}{\frac{\theta_{1}}{4}}\sin{\theta_{2}}e^{i(\epsilon_{1}-\epsilon_{2})}
&-\sin{\frac{\theta_{1}}{2}}\cos{\frac{\theta_{2}}{2}}e^{i\epsilon_{1}} \\
-\sin^{2}{\frac{\theta_{1}}{4}}\sin{\theta_{2}}e^{-i(\epsilon_{1}-\epsilon_{2})}
&1-2\sin^{2}{\frac{\theta_{1}}{4}}\sin^{2}{\frac{\theta_{2}}{2}}&
-\sin{\frac{\theta_{1}}{2}}\sin{\frac{\theta_{2}}{2}}e^{i\epsilon_{2}}\\
\sin{\frac{\theta_{1}}{2}}\cos{\frac{\theta_{2}}{2}}e^{-i\epsilon_{1}}&\sin{
\frac{\theta_{1}}{2}}\sin{\frac{\theta_{2}}{2}}e^{-i\epsilon_{2}}&\cos{\frac{
\theta_{1}}{2}}\\
 \end{array}\right).
\end{eqnarray}
%end{widetext}

\begin{figure}[!ht]
\centering
\includegraphics[height=3 in,width=3.5 in]{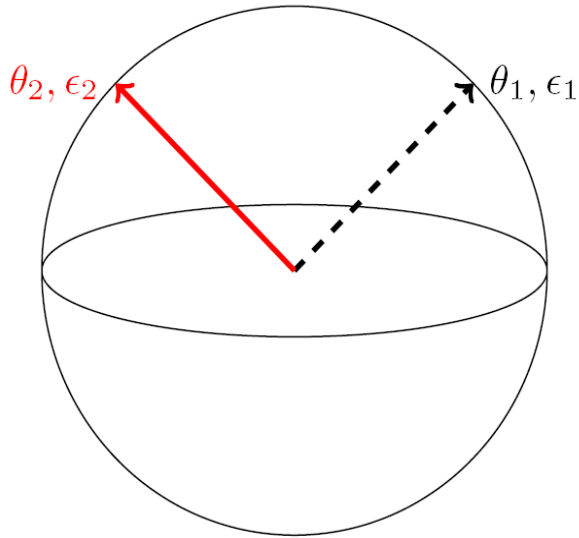}
\caption{The base manifold $U_{1}$ is characterized by two sets of angles
$0\leq\theta_{i}<\pi$, $0\leq\epsilon_{i}<2\pi$ which can be represented as two
vectors with polar angles $(\theta_{1},\epsilon_{1})$ and
$(\theta_{2},\epsilon_{2})$.}
\label{SU3_GP_Blochfig}
\end{figure}

In the above equation, the range on the angles $0\leq\theta_{i}<\pi$ and
$0\leq\epsilon_{i}<2\pi$ are chosen so that the absolute value of each
element of the time-evolution operator is positive \cite{aravind}. Hence $U_{1}$
can be represented as two vectors on a sphere, at angles
$(\theta_{1},\epsilon_{1})$ and $(\theta_{2},\epsilon_{2})$ respectively. This
is represented in Fig. (\ref{SU3_GP_Blochfig}).
Since the columns of a unitary operator correspond to normalized eigenvectors,
we can consider the last column of the matrix above,
$\vert\psi\rangle=(-\sin{\frac{\theta_{1}}{2}}\cos{\frac{\theta_{2}}{2}}e^{
i\epsilon_{1}},-\sin{\frac{\theta_{1}}{2}}\sin{\frac{\theta_{2}}{2}}e^{
i\epsilon_{2}},\cos{\frac{\theta_{1}}{2}})^{T}$, and evaluate the so-called
connection 1-form given by \cite{arno bohms book}
\begin{eqnarray}
\mathcal{A}=-i\langle\psi\vert d\vert\psi\rangle.
\end{eqnarray}
The Abelian geometric phase, given by $\gamma_{g}=\int{\mathcal{A}}$ is
evaluated to be
\begin{eqnarray}
\gamma_{g}=-\frac{1}{2}\int{\sin^{2}{\frac{\theta_{1}}{2}}\left((d\epsilon_{1}
+d\epsilon_{2})+\cos{\theta_{2}}(d\epsilon_{1}-d\epsilon_{2})\right)}.
\end{eqnarray}
If the various angles are relabelled $\epsilon_{1}\rightarrow-\gamma-\alpha$,
$\epsilon_{2}\rightarrow-\gamma+\alpha$, $\theta_{1}\rightarrow2\theta$ and
$\theta_{2}\rightarrow2\beta$, the formula above agrees with
\cite{byrd} and \cite{aravind}.
The time-evolution operator above can now be used as in the case of SU(2) to
evaluate the dynamic contribution $\int U^{\dagger}_{1}H(t)U_{1}$ and the
geometric contribution to the time evolution operator which is given by $-i\int
U^{\dagger}_{1}dU_{1}$, where $dU_{1}=\frac{dU_{1}}{d\theta_{i}}
d\theta_{i}+\frac{dU_{1}}{d\epsilon_{i}}d\epsilon_{i}$, $i=1,2$.

\begin{figure}[!ht]
\centering
\includegraphics[height=4 in,width=4 in]{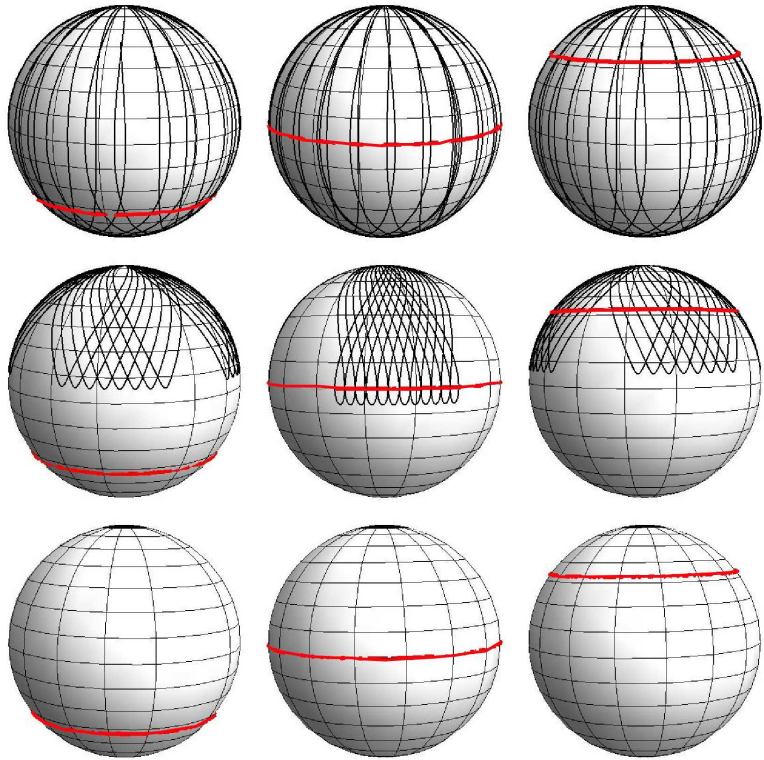}
\caption{ The base manifold corresponding to the results in
Fig. (\ref{Fig_kancheva}) for the three-level system of $\cite{kancheva}$. For
the first column,
$V_{1}=1$, $V_{2}=2$. The second column corresponds to $V_{1}=2$, $V_{2}=2$ and
the third to $V_{1}=2$, $V_{2}=1$. The rows correspond to $\delta=1$,
$\delta=5$
and $\delta=50$. The thin black curve describes  $(\theta_{1},\epsilon_{1})$
and
the thick red curve the set $(\theta_{2},\epsilon_{2})$.}
\label{Kancheva2vec}
\end{figure}

This description of the base manifold in terms of $(\theta_{i},\epsilon_{i})$
can now be used to describe the dynamics of various physical processes. Fig.
(\ref{Kancheva2vec}) represents the base manifold corresponding to the results
in Fig. (\ref{Fig_kancheva}). $(\theta_{1},\epsilon_{1})$ depend on all the
parameters that define the
system while $(\theta_{2},\epsilon_{2})$ depend only on the ratio $V_{1}/V_{2}$.
Also note that the maximum value of $\epsilon_{2}$, corresponding to the
maximum latitude traversed by the black curve, is inversely proportional to
$\delta$. Such observations can be used to control the dynamics of this
system.

\section{\label{conclusions}Conclusions}
The ability to decouple the time dependence of operator equations
from the non-commuting nature of the operators is the central
feature of unitary integration and also characterizes the Bloch sphere
representation for the evolution of a single spin.
By doing so, the quantum mechanical evolution is rendered a
``classical" picture of a rotating unit vector. For a two-level atom,
the Bloch sphere representation along with a phase completely
determines the time evolution operator. In this paper, we have
extended this program to deal with the time evolution operator belonging to
the SU(3) group. This complements the work in \cite{newui} for SU(4)
Hamiltonians of two qubit systems. We have also extended the
analysis of geometric phase to three-level systems by providing an explicit
coordinate representation for the SU(3) time evolution operator.
\appendix
\section{\label{appendix}Alternative derivations for a general SU(3)
Hamiltonian.}
 Consider a three-level Hamiltonian written in terms of the Gell-Mann
matrices \cite{georgi} as $H(t)=\sum_{i=1}^{8}a_{i}\bm{\lambda}_{i}$.  To
exploit the fact that this Hamiltonian is a subgroup
of four-level problems, it is represented in terms of the O
matrices  \cite{restui} as
\begin{eqnarray}\label{SU(3) derivation}
\fl
2\frac{a_{8}}{\sqrt{3}}\mathbf{O}_{2}+(a_{3}-\frac{a_{8}}{\sqrt{3}})\mathbf
{O}_{3}+(2a_{3}+2\frac{a_{8}}{\sqrt{3}})\mathbf{O}_{4}
+a_{4}\mathbf{O}_{5}+a_{5}\mathbf{O}_{6}+2a_{4}\mathbf{O}_{7}+2a_{5}\mathbf{O}_{
8}+\nonumber\\
\fl\qquad
a_{1}\mathbf{O}_{9}+
a_{2}\mathbf{O}_{10}+2a_{1}\mathbf{O}_{11}+2a_{2}\mathbf{O}_{12}+2a_{6}\mathbf{O
}_{13}+2a_{6}\mathbf{O}_{14}-2a_{7}\mathbf{O}_{15}+2a_{7}\mathbf{O}_{16}.
\end{eqnarray}
This embeds the Hamiltonian $H(t)=\sum_{i}a_{i}\bm{\lambda}_{i}$ as a
4$\times$4 matrix with zeros along the last row and column. In such a
representation,
the various entries of the Hamiltonian equation (\ref{Hamiltonian}) are given
by
\begin{eqnarray}
H^{(4-2)}=\frac{1}{\sqrt{3}}a_{8}\mathbf{I}^{(2)}+a_{1}\bm{\sigma}_{1}+a_{2}
\bm{\sigma}_{2}+a_{3}\bm{\sigma}_{3},\\
H^{(2)}=-\frac{1}{\sqrt{3}}a_{8}\mathbf{I}^{(2)}-\frac{1}{\sqrt{3}}a_{8}\bm{
\sigma}_{1},\\
\mathbf{V}=\frac{1}{2}(a_{4}-ia_{5})\mathbf{I}^{(2)}+\frac{1}{2}(a_{6}-ia_{7}
)\bm{\sigma}_{1}\\\nonumber
-i\frac{1}{2}(a_{6}-ia_{7})\bm{\sigma}_{2}+\frac{1}{2}(a_{4}-ia_{5})\bm{
\sigma}_{3}.
\end{eqnarray}
Writing $\mathbf{z}$ in the standard Clifford basis as
$\mathbf{z}=\frac{1}{2}z_{4}\bm{I}^{(2)}-\frac{i}{2}\sum_{i}z_{i}\bm{\sigma}_{i}
$, it follows from equation~(\ref{zdot}) that $z_{1}=iz_{2}$ and $z_{3}=iz_{4}$
and
the equation reduces precisely to equation (\ref{su3_z_eqn}).
 The geometry described in Section \ref{su3} can thus be derived from either of
these decompositions of the time evolution operator.

The SU(3) subgroup in equation (\ref{SU(3) derivation}) is one among many SU(3)
subgroups embedded in SU(4). Another choice corresponds to the
Dzyaloshinskii-Moriya interaction Hamiltonian \cite{dz1,dz2} and is also of
interest because the 4$\times$4 matrices now do not have a trivial row and
column of zeros.  In the
two-spin basis, this Hamiltonian is given by
\begin{eqnarray}\label{expandedDV}
\fl
H(t)=\sum_{i}c_{i}\mathbf{O}_{i}=a_{1}(\mathbf{O}_{2}+\mathbf{O}_{3})+2
a_{2}(\mathbf{O}_{15}+\mathbf{O}_{16})+2 a_{3}(\mathbf{O}_{14}\--
\mathbf{O}_{13})+2 a_{4}(\mathbf{O}_{7}+\mathbf{O}_{11})\nonumber\\
\fl\qquad
+a_{5}(\mathbf{O}_{6}+\mathbf{O}_{10})+a_{6}(\mathbf{O}_{5}+\mathbf{O}_{9})+2
a_{7}(\mathbf{O}_{8}+\mathbf{O}_{12})+\frac{2 a_{8}}{\sqrt{3}}(2
\mathbf{O}_{4}\--\mathbf{O}_{13}\--\mathbf{O}_{14}).
\end{eqnarray}
The correspondence between the coefficients in terms of $\mathbf{O}$ and in
terms of the $\bm{\lambda}$ matrices is : $c_{1}=0$, $c_{2}=a_{1}$,
$c_{3}=a_{1}$,
$c_{4}=4a_{8}/\sqrt{3}$, $c_{5}=a_{6}$, $c_{6}=a_{5}$, $c_{7}=2a_{4}$,
$c_{8}=2a_{7}$,  $c_{9}=a_{6}$, $c_{10}=a_{5}$, $c_{11}=2a_{4}$,
$c_{12}=2a_{7}$, $c_{13}=-2a_{3}-2a_{8}/\sqrt{3}$,
$c_{14}=2a_{3}-2a_{8}/\sqrt{3}$, $c_{15}=2a_{2}$ and $c_{16}=2a_{2}$. Relabeling
of the states $1\rightarrow2$, $2\rightarrow3$,
$3\rightarrow4$ and $4\rightarrow1$ expresses the Hamiltonian as
\begin{eqnarray}
H^{(4-2)}=\frac{1}{\sqrt{3}}a_{8}\mathbf{I}^{(2)}-a_{3}\bm{\sigma}_{1}-a_{2}
\bm{\sigma}_{2}-a_{1}\bm{\sigma}_{3},\\
H^{(2)}=-\frac{1}{\sqrt{3}}a_{8}\mathbf{I}^{(2)}-\frac{1}{\sqrt{3}}a_{8}\bm{
\sigma}_{1},\\
\mathbf{V}=\frac{1}{2}(a_{6}-ia_{7})\mathbf{I}^{(2)}+\frac{1}{2}(a_{6}-ia_{7}
)\bm{\sigma}_{1}\\\nonumber
-\frac{1}{2}(a_{5}+ia_{4})\bm{\sigma}_{2}-\frac{1}{2}(a_{4}-ia_{5})\bm{
\sigma}_{3}.
\end{eqnarray}
If $\mathbf{z}$ is written in terms of the standard Clifford basis
$(\hat{\mathbf{I}},-i\vec{\bm{\sigma}})$ as
$\mathbf{z}=\frac{1}{2}z_{4}\mathbf{I}^{(2)}-\frac{i}{2}\sum_{i=1}^{3}z_{i}\bm{
\sigma}_{i} $ ,
 it follows from equation (\ref{zdot}) that
$z_{1}=i z_{4}$ and
$z_{2}=i z_{3}$. This is consistent with the
parameter count that since the inhomogeneity $\mathbf{V}$ has only
two free complex parameters (namely $V_{1}=a_{6}-ia_{7}$ and
$V_{2}=a_{4}-ia_{5}$), the complex $\mathbf{z}$ matrix
should be composed only of two independent complex parameters,
$z_{1}$ and $z_{2}$. With the above analysis, equation~(\ref{zdot}) becomes for
the
pair of complex numbers
\begin{equation}
\frac{1}{2}\dot{z}_{\mu}=\frac{1}{2}X_{\mu}-iF_{\mu\nu}z_{\nu}+2G_{\nu}z_{\nu}z_
{ \mu}
;\;\mu,\nu=1,2.
\end{equation}
Here $X=(V_{1}/2,-i V_{2}/2)$,
$G=(2V_{1}^{*},2i V_{2}^{*})$ and
\begin{equation*}
-i F=\left(\begin{array}{cc}
i a_{3}-\sqrt{3} i a_{8}&a_{1}+i a_{2}\\
-a_{1}+i a_{2}&-i a_{3}-\sqrt{3}i a_{8}\\
\end{array} \right).
\end{equation*}

Paralleling the technique employed to solve an
SO(5) Hamiltonian in \cite{newui,newui2}, we transform $\mathbf{z}$ into a
complex
vector $\vec{m}$: $m_{\mu}=\frac{-2z_{\mu}e^{i\phi}}{D}$ and
$m_{3}=\frac{e^{i\phi}}{D}$ such that $\vert m_{1} \vert^{2}+\vert
m_{2} \vert^{2}+\vert m_{3} \vert^{2}=1$,with
$D=(1+4(\vert z_{1} \vert^{2}+\vert z_{2}
\vert^{2}))^{1/2}$. This leads to the new set of
evolution equations
\begin{small}
\begin{equation}
\dot{\vec{m}}=\left(\begin{array}{ccc}
i a_{3}-\sqrt{3} i a_{8}&a_{1}+i a_{2}&-a_{6}+i a_{7}\\
-a_{1}+i a_{2}&-i a_{3}-\sqrt{3}i a_{8}&a_{5}+i a_{4}\\
a_{6}+i a_{7}&-a_{5}+i a_{4}&0
\end{array} \right)\vec{m}.
\end{equation}
\end{small}
This can be written as an equation describing the rotation of the real and
imaginary components of the vector
$\vec{m}=(m_{1r},m_{2r},m_{3r},m_{1i},m_{2i},m_{3i})^{T}$,
\begin{eqnarray}\label{rotation_SU4}
\fl
\dot{\vec{m}}=\left(\begin{array}{cccccc}
0&a_{1}&-a_{6}&-a_{3}+\sqrt{3}a_{8}&-a_{2}&-a_{7}\\
-a_{1}&0&a_{5}&-a_{2}&a_{3}+\sqrt{3}a_{8}&-a_{4}\\
a_{6}&-a_{5}&0&-a_{7}&-a_{4}&0\\
a_{3}-\sqrt{3}a_{8}&a_{2}&a_{7}&0&a_{1}&-a_{6}\\
a_{2}&-a_{3}-\sqrt{3}a_{8}&a_{4}&-a_{1}&0&a_{5}\\
a_{7}&a_{4}&0&a_{6}&-a_{5}&0\\
\end{array} \right)\vec{m}.
\end{eqnarray}
Here, the coefficients $c_{i}$ are written in terms of the
coefficients $a_{i}$, whose correspondence was given earlier in
this section. Also note that $m_{\mu}=m_{\mu
r}+im_{\mu i}$, $D=(1+\vert z_{1}\vert^{2}+\vert
z_{2}\vert^{2})^{\frac{1}{2}}$ and
$\dot{\phi}=(V^{*}_{\nu}z_{\nu}+V_{\nu}z^{*}_{\nu})$.
Simplifying this leads to the equation
$i\dot{\phi}=-2(X_{\mu}z_{\mu}^{*}-X_{\mu}^{*}z_{\mu})$
for the evolution of $\phi$ which is clearly real but determined only to within
a constant.
A little algebra yields for the effective
Hamiltonian given by equation~(\ref{heffup}),
\begin{small}
\begin{eqnarray*}
H^{(4-2)}-\frac{1}{(D+1)}(\mathbf{z}\mathbf{V}^{\dagger}+\mathbf{V}\mathbf{z}^{
\dagger})-
\frac{1}{2(D+1)^{2}}(\mathbf{z}\mathbf{V}^{\dagger}\mathbf{z}\mathbf{z}^{\dagger
} \- +\mathbf{z}\mathbf{z}^{\dagger}\mathbf{V}\mathbf{z}^{\dagger}),
\end{eqnarray*}
\end{small}
and for the effective Hamiltonian given by equation~(\ref{heffdown}), the
expression
$H^{(2)}+(\mathbf{z}^{\dagger}\mathbf{V}+\mathbf{V}^{\dagger}\mathbf{z})/2.$

Another representation of the SU(3) subgroup of SU(4) Hamiltonians is given by
the so called ``Pl\"{u}cker coordinate'' representation of the SU(4)
group discussed
in
\cite{newui,newui2}. For an arbitrary SU(4) matrix, the Pl\"{u}cker coordinates
are
defined as a set of six parameters $(P_{12},P_{13},P_{14},P_{23},P_{24},P_{34})$
such that $P_{12}P_{34}-P_{13}P_{24}+P_{14}P_{23}=0$ and $\sum \vert
P_{ij}\vert^{2}=1$. They can be written in terms of the unit vector $\vec{m}$
and are given by
\begin{equation}
\left(\begin{array}{c}
P_{12}\\
P_{13}\\
P_{14}\\
P_{23}\\
P_{24}\\
P_{34}\\
\end{array} \right)=\frac{1}{2}
\left(\begin{array}{c}
im_{6}-m_{5}\\
im_{1}+m_{2}\\
-im_{3}+m_{4}\\
-im_{3}-m_{4}\\
-im_{1}+m_{2}\\
im_{6}+m_{5}\\
\end{array} \right).
\end{equation}
The linear equation of motion for $\vec{m}$ translates into an evolution
equation for $\mathbf{P}=(P_{12},-P_{13},P_{14},P_{23},P_{24},P_{34})$ of the
form $i\dot{\mathbf{P}}=\mathbf{H}_{P}\mathbf{P}$. Here, $\mathbf{H}_{P}$ is
given by
\begin{equation}
 \mathbf{H}_{P}=\left(\begin{array}{cc}
 \mathbf{H}_{P1}&\mathbf{V}_{P}\\
 \mathbf{V}^{\dagger}_{P}&\mathbf{H}_{P2}\\
                \end{array}\right),
\end{equation}
where
\begin{eqnarray*}
\fl
%%Upper Block of Plucker  Hamiltonian
 \mathbf{H}_{P1}=\left(\begin{array}{ccc}
2a_{8}/\sqrt{3}&a_{64-}+ia_{75-}&a_{64-}+ia_{75-}\\
a_{64-}-ia_{75-}&-a_{1}&a_{8}/\sqrt{3}\\
a_{64-}-ia_{75-}&a_{8}/\sqrt{3}&-a_{1}
                        \end{array}
\right),\\
\fl
%%Lower Block of Plucker  Hamiltonian
 \mathbf{H}_{P2}=\left(\begin{array}{ccc}
a_{1}&-a_{8}/\sqrt{3}&-a_{64-}-ia_{75-}\\
-a_{8}/\sqrt{3}&a_{1}&-a_{64-}-ia_{75-}\\
-a_{64-}+ia_{75-}&-a_{64-}+ia_{75-}&-2a_{8}/\sqrt{3}
                        \end{array}
\right),\\
\fl
%%Off diagonal Block of Plucker Hamiltonian
 \mathbf{V}_{P}=\left(\begin{array}{ccc}
-a_{64+}-ia_{75+}&a_{64+}+ia_{75+}&0\\
a_{32-}&0&-a_{64+}-ia_{75+}\\
0&-a_{32-}&a_{64+}-ia_{75+}
                        \end{array}
\right).
\end{eqnarray*}
In the above equation, $a_{ij\pm}$ denotes $a_{i}\pm a_{j}$.

\end{document}